\documentclass[epj]{webofc}
\usepackage[varg]{txfonts}   
\wocname{EPJ Web of Conferences}
\woctitle{ICNFP 2017}
\def\Acknowledgements{\bigskip  \bigskip \begin{center} \begin{large}
             \bf ACKNOWLEDGEMENTS \end{large} \end{center} }
\newcommand{\Pp}{\ensuremath{{p}}}

\newcommand{\PH}{\ensuremath{{H}}}
\newcommand{\JPsi}{\ensuremath{{J}\hspace{-.08em}/\hspace{-.14em}\psi}\xspace}
\providecommand{\Z}{\ensuremath{\mathrm{Z}}\xspace}

\newcommand{\mh}{\ensuremath{m_\PH}\xspace}
\newcommand{\BF}{\ensuremath{\mathcal{B}}\xspace}
\newcommand{\Pem}{\ensuremath{\mathrm{e}^-}}
\newcommand{\Pep}{\ensuremath{\mathrm{e}^+}}

\newcommand{\mumu}{\ensuremath{\Pgmp\Pgmm}\xspace}
\newcommand{\ee}{\ensuremath{\Pep\Pem}\xspace}
\newcommand{\Pgmm}{\ensuremath{{\mu^-}}}
\newcommand{\Pgmp}{\ensuremath{{\mu^+}}}
\newcommand{\tautau}{\ensuremath{\tau^{+}\tau^{-}}\xspace}
\newcommand{\hmm}{\ensuremath{\PH\to\mumu}\xspace}
\newcommand{\hee}{\ensuremath{\PH\to\ee}\xspace}
\newcommand{\htautau}{\ensuremath{\PH\to\tautau}\xspace}

\newcommand{\bbbar}{\ensuremath{{b\overline{b}}}\xspace}
\newcommand{\brinv}{\ensuremath{\mathcal{B}(\PH\to \text{inv})}\xspace}
\newcommand{\higgsbrobs}{\ensuremath{0.24}}

\newcommand{\higgsbrexp}{\ensuremath{0.23}}

\newcommand{\Et}{\ensuremath{E_\mathrm{T}}}
\newcommand{\met}{\ensuremath{\Et^{\mathrm{miss}}}}

\newcommand{\Pgm}{\ensuremath{{\mu}}}
\newcommand{\Pgt}{\ensuremath{{\tau}}}
\newcommand{\Pe}{\ensuremath{{e}}}
\providecommand{\tauh}{\ensuremath{\Pgt_\mathrm{h}}\xspace}

\newcommand{\Hb}{\ensuremath{{\mathrm{h}}}\xspace}
\newcommand{\ab}{\ensuremath{\mathrm{a}}\xspace}

\providecommand{\PQb}{\ensuremath{b}\xspace} 
\newcommand{\processmmbb}{\ensuremath{{\Hb}\to{\ab\ab}\to2\Pgm2{\PQb}}\xspace}
\newcommand{\processmmtt}{\ensuremath{{\Hb}\to{\ab\ab}\to2\Pgm2\Pgt}\xspace}
\newcommand{\processtttt}{\ensuremath{{\Hb}\to{\ab\ab}\to4\Pgt}\xspace}
\newcommand{\processmmmm}{\ensuremath{{\Hb}\to{\ab\ab}\to4\Pgm}\xspace}

\newcommand{\fb}{\mathrm{fb}}
\newcommand{\fbinv}{\mathrm{fb}^{-1}}
\newcommand{\GeV}{\mathrm{{GeV}}}
\newcommand{\TeV}{\mathrm{{TeV}}}

\begin{document}
\selectlanguage{english}
\title{Search for rare and exotic Higgs Boson decay modes}
%
%

\author{Junquan Tao\inst{1}\fnsep\thanks{\email{taojq@mail.ihep.ac.cn}} on behalf of the CMS collaboration
}

\institute{Institute of High Energy Physics, Chinese Academy of Sciences, 100049 Beijing, China
}

\abstract{%
   The latest results in the search for rare and exotic Higgs boson decays in proton-proton collision events collected with the CMS detector at the LHC are presented. The searches are performed for several decay modes of Higgs boson including $\mathrm{H}\rightarrow{\rm X (X \rightarrow2\ell)\gamma}$ ($X= {\rm Z}, \gamma^* $ and $\ell={\rm e},\mu$), $\mathrm{H}\rightarrow{ \mu\mu / {\rm e}{\rm e}}$, invisible decays, lepton flavour violating decays and Higgs decay to light scalars or pseudo-scalars. No hint for new physics has been found from the analyzed results with the full LHC run-1 data collected during 2011 and 2012 at $\sqrt{s}=7-8~\TeV$ and with the run-2 data at $\sqrt{s}=13~\TeV$ collected during 2015 and 2016. Limits are set for all the searches which have been performed by CMS.
}
\maketitle
\section{Introduction}
\label{intro}

The discovery of the Higgs boson with a mass of 125.09 $\pm$ 0.24 $\GeV$~\cite{Aad:2012tfa,Chatrchyan:2012xdj,Chatrchyan:2013lba} at the Large Hadron Collider (LHC) has generated great interest in exploring its properties.
Many measurements from LHC have confirmed that the Higgs boson has properties, including spin,
CP, and coupling strengths, that are compatible with those expected for
the Higgs boson of the SM~\cite{Englert:1964et,Higgs:1964pj,Guralnik:1964eu}. However the scalar sector is not well known experimentally yet, and current
measurements could still accommodate for large contributions of new physics in this sector.
Constraints on new physics from the LHC are still relatively loose~\cite{Khachatryan:2016vau}, which leaves room for beyond the standard model (BSM) physics.
Many Higgs rare decays of Higgs boson in SM have yet to be observed, which may be sensitive to new physics if additional Higgs couplings exist.
Observing exotic decays of the Higgs boson would be a striking direct evidence for the existence of physics beyond the SM.
The latest results in the search for rare and exotic Higgs boson decays in proton-proton collision events collected with the CMS detector at the LHC will be summarized in this paper. The searches are performed for an extensive set of decay modes of Higgs boson including $\mathrm{H}\rightarrow{\rm X (X \rightarrow2\ell)\gamma}$ ($X= {\rm Z}, \gamma^* $ and $\ell={\rm e},\mu$), $\mathrm{H}\rightarrow{ \mu\mu / {\rm e}{\rm e}}$, invisible decays, lepton flavour violating decays and Higgs decay to light scalars or pseudo-scalars, with the full LHC run-1 data collected during 2011 and 2012 at $\sqrt{s}=7-8~\TeV$ and with the run-2 data at $\sqrt{s}=13~\TeV$ collected during 2015 and 2016.
The searches are generally performed in a model independent approach, and exclusion limits
in terms of production cross section times the corresponding decay branching ratio are presented.
Many of the results are interpreted according to beyond-SM (BSM) Higgs scenarios, which include the
Two Higgs Doublet Model (2HDM) and the Singlet Model.

\section{$\mathrm{H}\rightarrow{\rm X \gamma}$}
\label{HXg}

\begin{figure}[h]
 \begin{center}
   \includegraphics[width=0.48\textwidth]{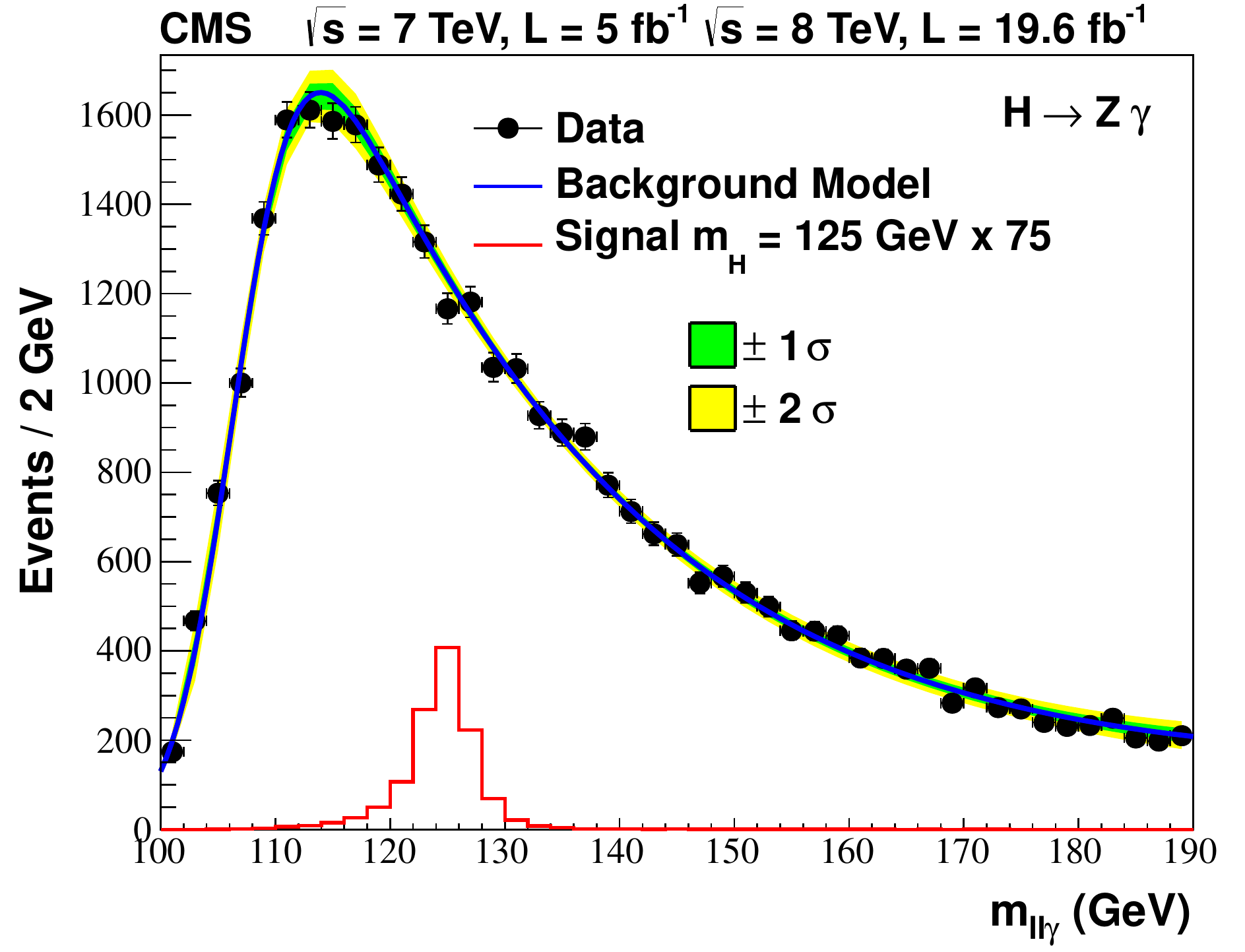}
   \includegraphics[width=0.51\textwidth]{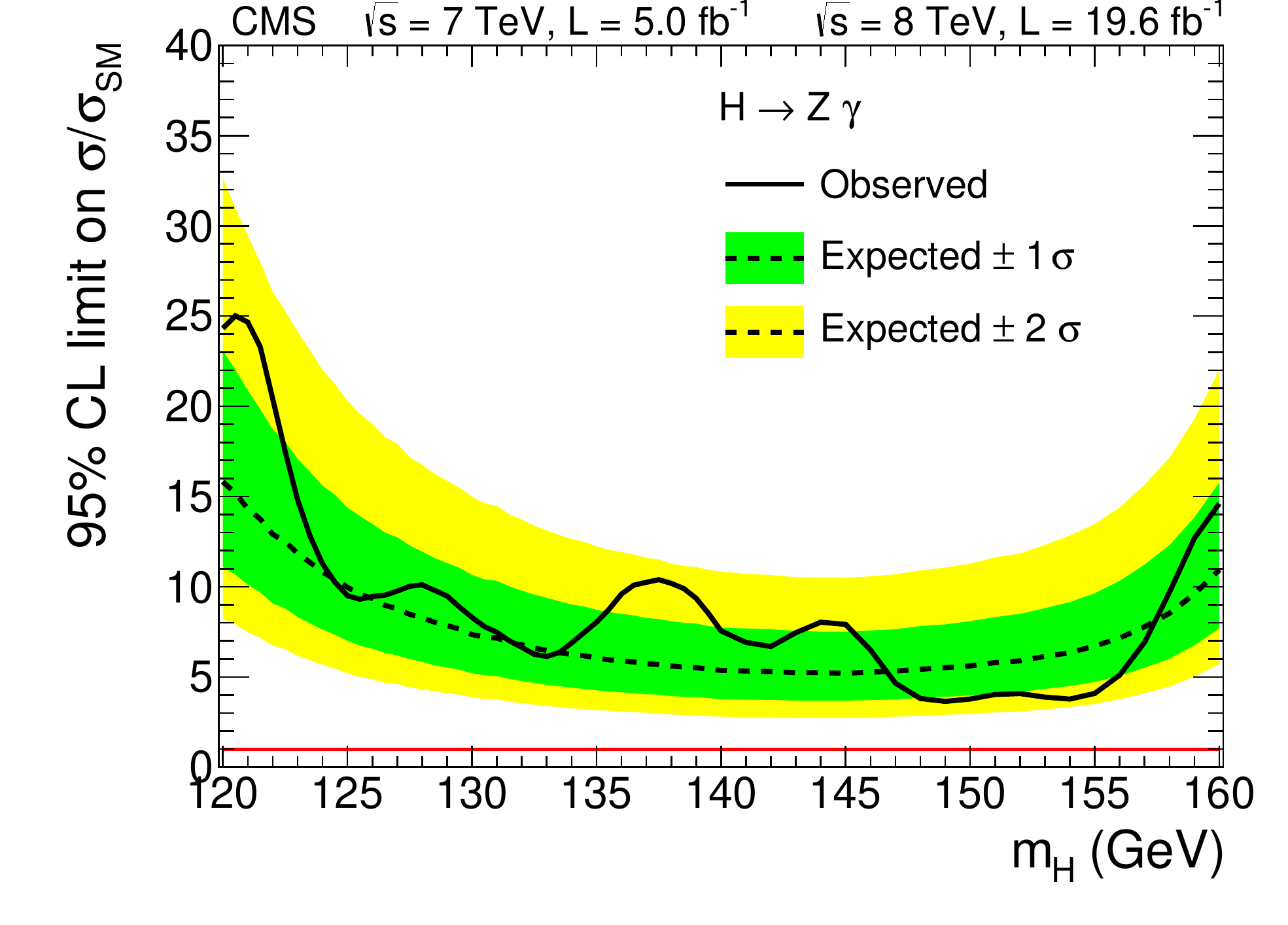}
   \caption{Left plot shows the $m_{\ell\ell\gamma}$ spectrum in the electron and the muon
channels for the 7 and 8 $\TeV$ data combined, without weighting by the expected
signal to background ratio of the individual data samples. Also shown is the expected
signal due to a 125 $\GeV$ standard model Higgs boson, scaled by 75,  and the sum of the individual
fits made to the data for each channel and event class. The uncertainty band reflects the statistical uncertainty from the fits to the data.
The exclusion limit on the cross section times the branching fraction of a Higgs boson decaying
into a $\Z$ boson and a photon divided by the SM value is shown in the right plot.
The red line represents the same cross section times the branching fraction as the SM prediction~\cite{Chatrchyan:2013vaa}.}
  \label{fig:HZg}
 \end{center}
\end{figure}

Within the SM, the partial width for the $\PH\to\Z\gamma$ decay channel ($\Gamma_{\Z\gamma}$)
is rather small, resulting in a branching fraction between  0.11\% and 0.25\% in  the
120 $-$ 160 $\GeV$~\cite{Cahn:1978nz,Bergstrom:1985hp} mass range.
A measurement of $\Gamma_{\Z\gamma}$ provides important information
on the underlying dynamics of the Higgs sector
because it is induced by loops of heavy charged particles, just as for the
$\PH\to\gamma\gamma$ decay channel. This paper summarizes the most recent search for a Higgs boson in the $\PH\rightarrow\Z\gamma$ final state
at the LHC in the 120 $-$ 160 $\GeV$ mass range, with the $\Z$ boson decaying into an electron or a muon pair~\cite{Chatrchyan:2013vaa}.
This is a clean final-state topology with an effective mass peak  resolution of about 1-3\%.
Events were collected at center-of-mass energies of 7 $\TeV$ and 8 $\TeV$, corresponding to integrated luminosities of
5.0 $\fbinv$ and 19.6 $\fbinv$, respectively.
The selected events are required to have
opposite-sign electron or muon pairs. The mass spectrum for all channels combined is shown in the left plot of Fig.~\ref{fig:HZg}.
No excess above standard model predictions has been found  in the
120--160 $\GeV$ mass range
and the first limits on the Higgs boson production
cross section times the $\PH\to\Z\gamma$  branching fraction at the LHC have been derived,
as shown in the right plot of Fig.~\ref{fig:HZg}.
For a standard model Higgs boson mass of 125 $\GeV$ the expected limit at the 95$\%$ confidence level is 10 and the observed limit is 9.5.

\begin{figure}[h]
 \begin{center}
   \includegraphics[width=0.49\textwidth]{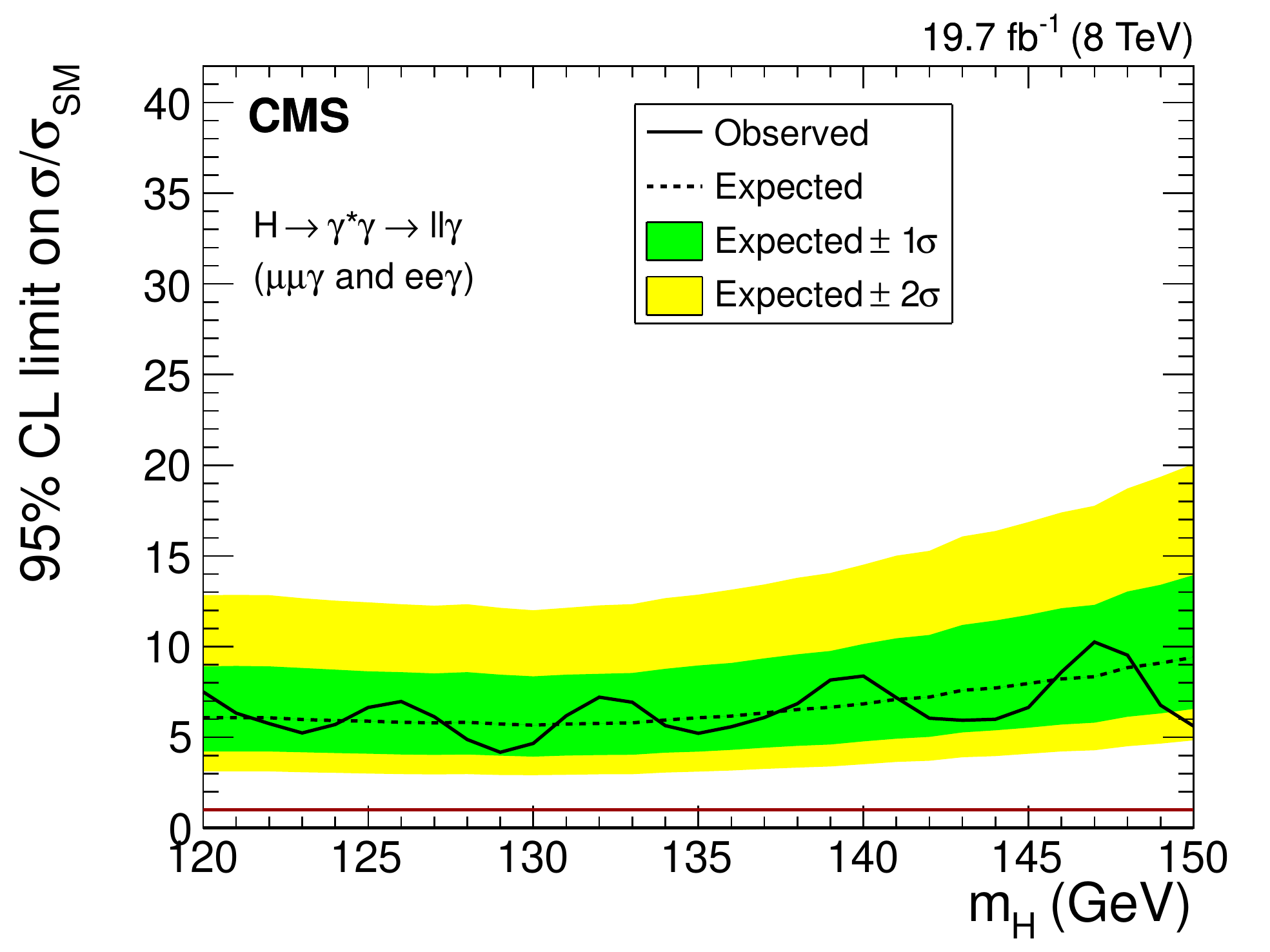}
   \includegraphics[width=0.49\textwidth]{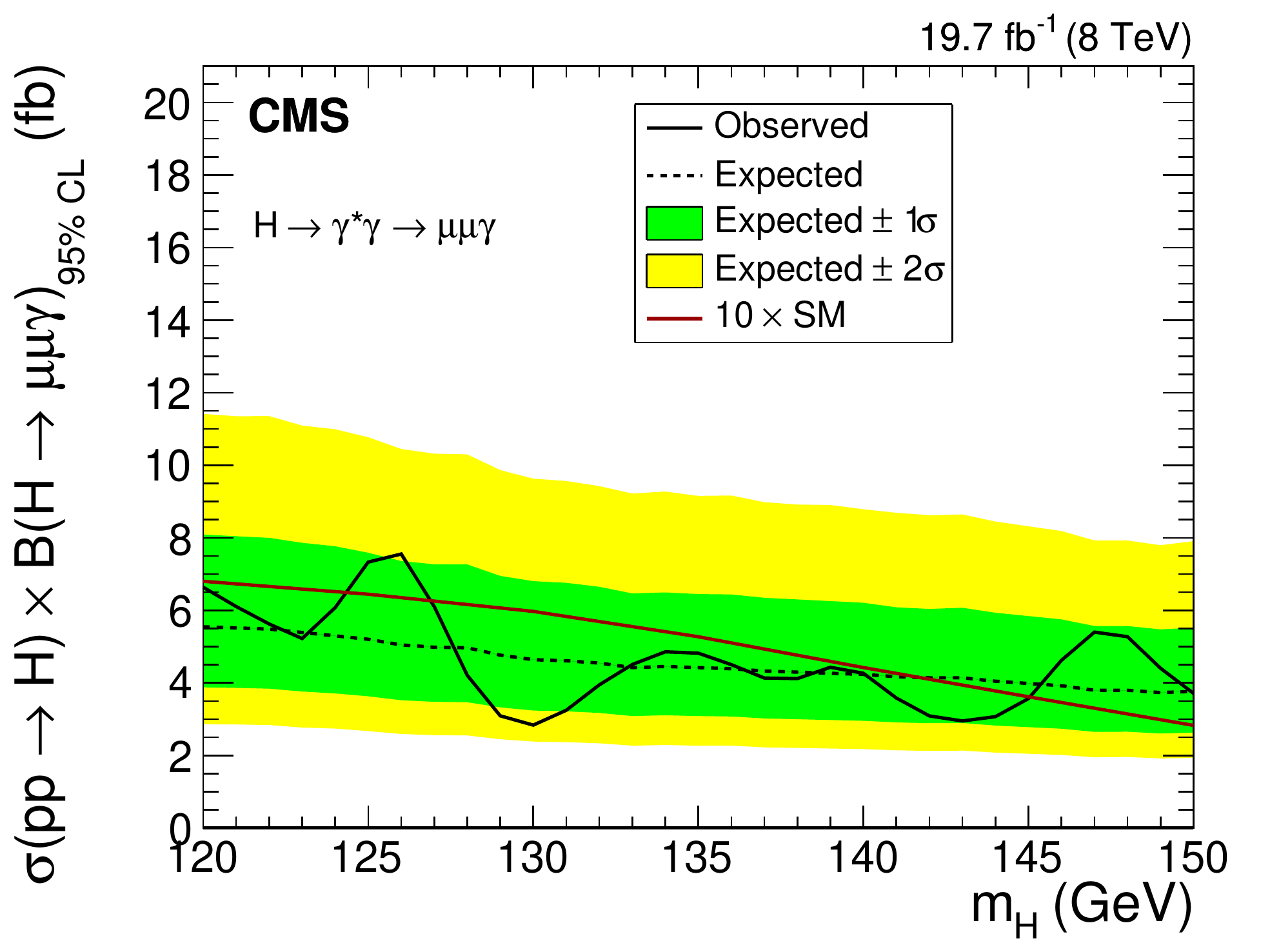}
   \caption{(Left) The
    95\% CL exclusion limit, as a function of the mass hypothesis,
    $m_\PH$, on $\sigma/\sigma_{SM}$, the cross section times the
    branching fraction of a Higgs boson decaying into a photon and a
    lepton pair with $m_{\ell\ell} < 20\GeV$, divided by the SM value.
    (Right) The 95\% CL exclusion limit on $\sigma(\Pp\Pp\to \PH)\,\mathcal{B}(\PH\to\mu\mu\gamma)$,
      with $m_{\mu\mu} < 20\GeV$, for a Higgs-like particle, as a function of the mass hypothesis, $m_\PH$~\cite{Khachatryan:2015lga}.}
  \label{fig:Hgg}
 \end{center}
\end{figure}

The rare decay into the $\ell\ell\gamma$ final state of the Higgs boson is a rich source
of information that can enhance our understanding of its basic properties and probe novel
couplings predicted by extensions of the standard model of particle physics.
The search for a Higgs boson decay $\PH\to\gamma^*\gamma\to\ell\ell\gamma$ is performed using
proton-proton collision data recorded with the CMS detector at the LHC at a centre-of-mass
energy of 8 $\TeV$, corresponding to an integrated luminosity of 19.7 $\fbinv$~\cite{Khachatryan:2015lga}. No
excess above the background predictions has been found in the three-body invariant mass
range $120<m_{\ell\ell\gamma}<150~\GeV$. Limits on the Higgs boson production cross section
times the $\PH\to\gamma^*\gamma\to\ell\ell\gamma$ branching fraction divided by the SM
values have been derived, as shown in the left plot of Fig.~\ref{fig:Hgg}.
The observed limit for $m_\PH=125~\GeV$ is about 6.7 times the
SM prediction.  Limits at 95\% CL on
$\sigma(\Pp\Pp\to \PH)\,\mathcal{B}(\PH\to\mu\mu\gamma)$ for a narrow resonance are also
obtained in the muon channel, as shown in the right plot of Fig.~\ref{fig:Hgg}.  The observed limit for $m_\PH = 125~\GeV$ is 7.3 $\fb$.
In addition, a search is performed for $\PH \to \JPsi\gamma\to\mu\mu\gamma$ decay for
$m_\PH=125~\GeV$, which is sensitive to the Higgs boson coupling to charm quark and a
promising way to access the couplings of the Higgs boson to the second generation quarks
at the LHC~\cite{Khachatryan:2015lga}.
Events consistent with the $\JPsi$ in dimuon invariant mass are used to set a 95\% CL
limit on the branching fraction $\mathcal{B}(\PH\to\JPsi\gamma) < 1.5\times10^{-3}$,
that is, 540 times the SM prediction for $m_\PH=125~\GeV$.

\section{$\mathrm{H}\rightarrow{ \mu\mu / {\rm e}{\rm e}}$}
\label{Hll}

\begin{figure}[!hbtp]
  \centering
   \includegraphics[width=0.325\textwidth]{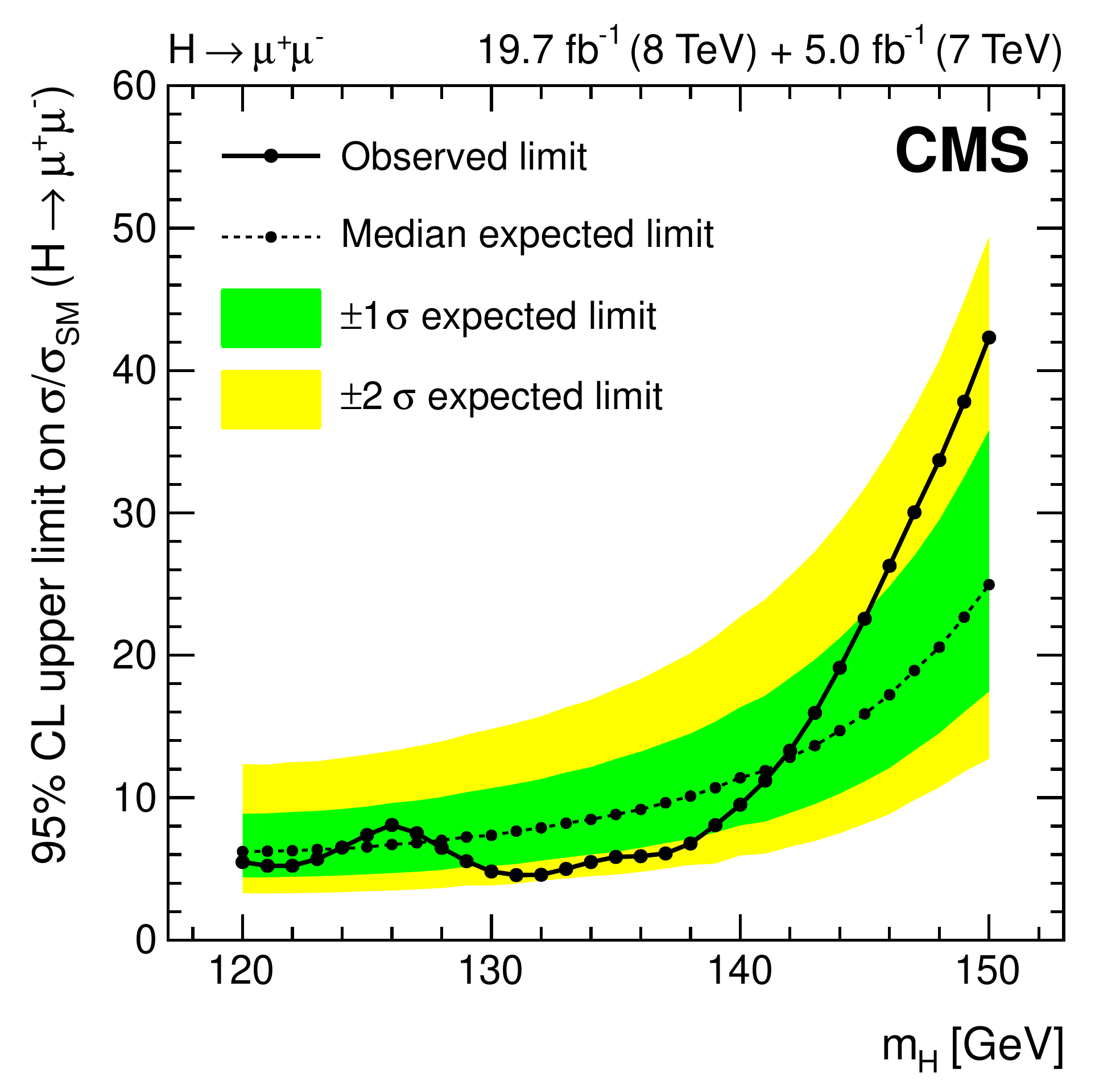}
   \includegraphics[width=0.325\textwidth]{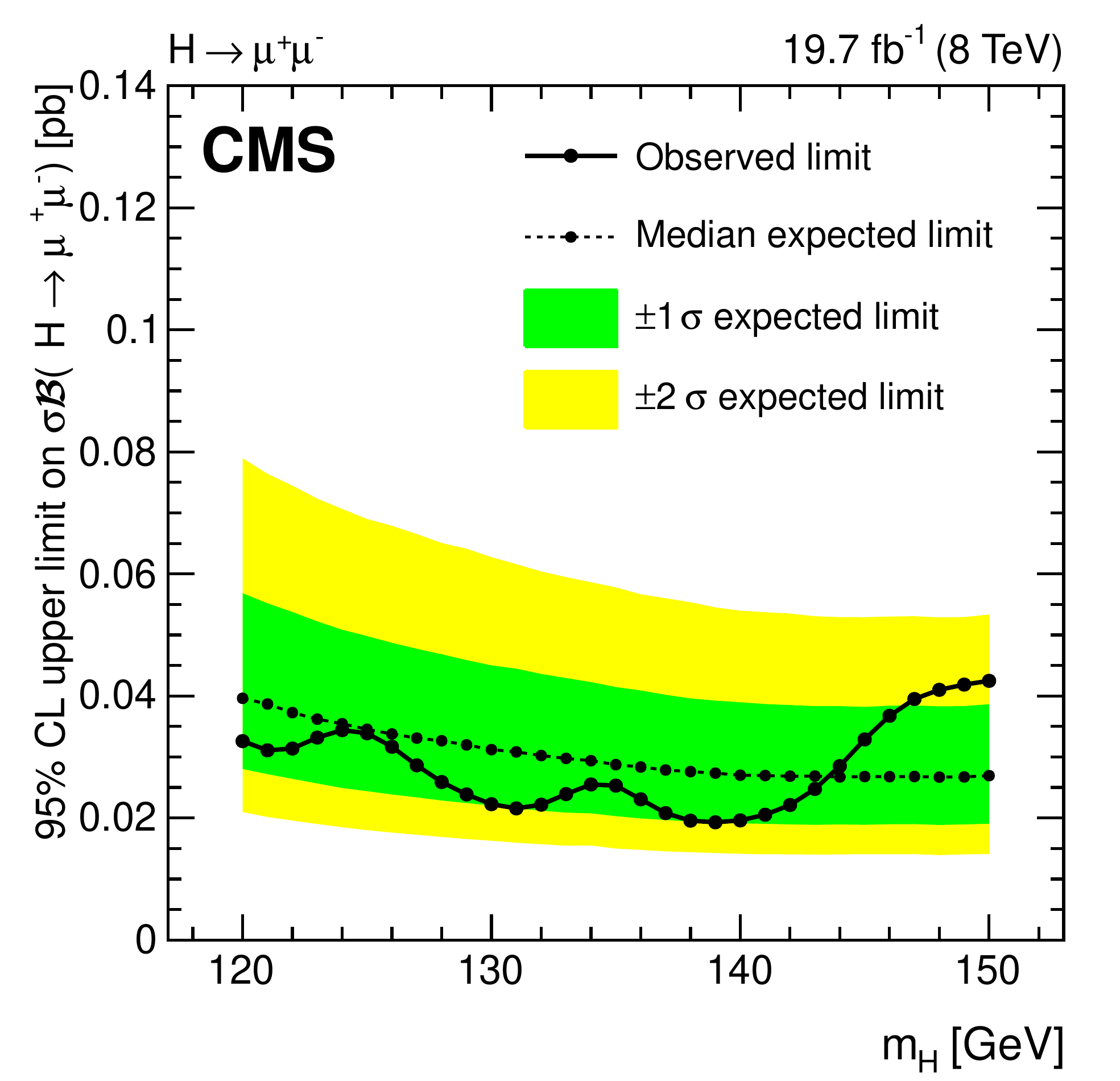}
    \includegraphics[width=0.325\textwidth]{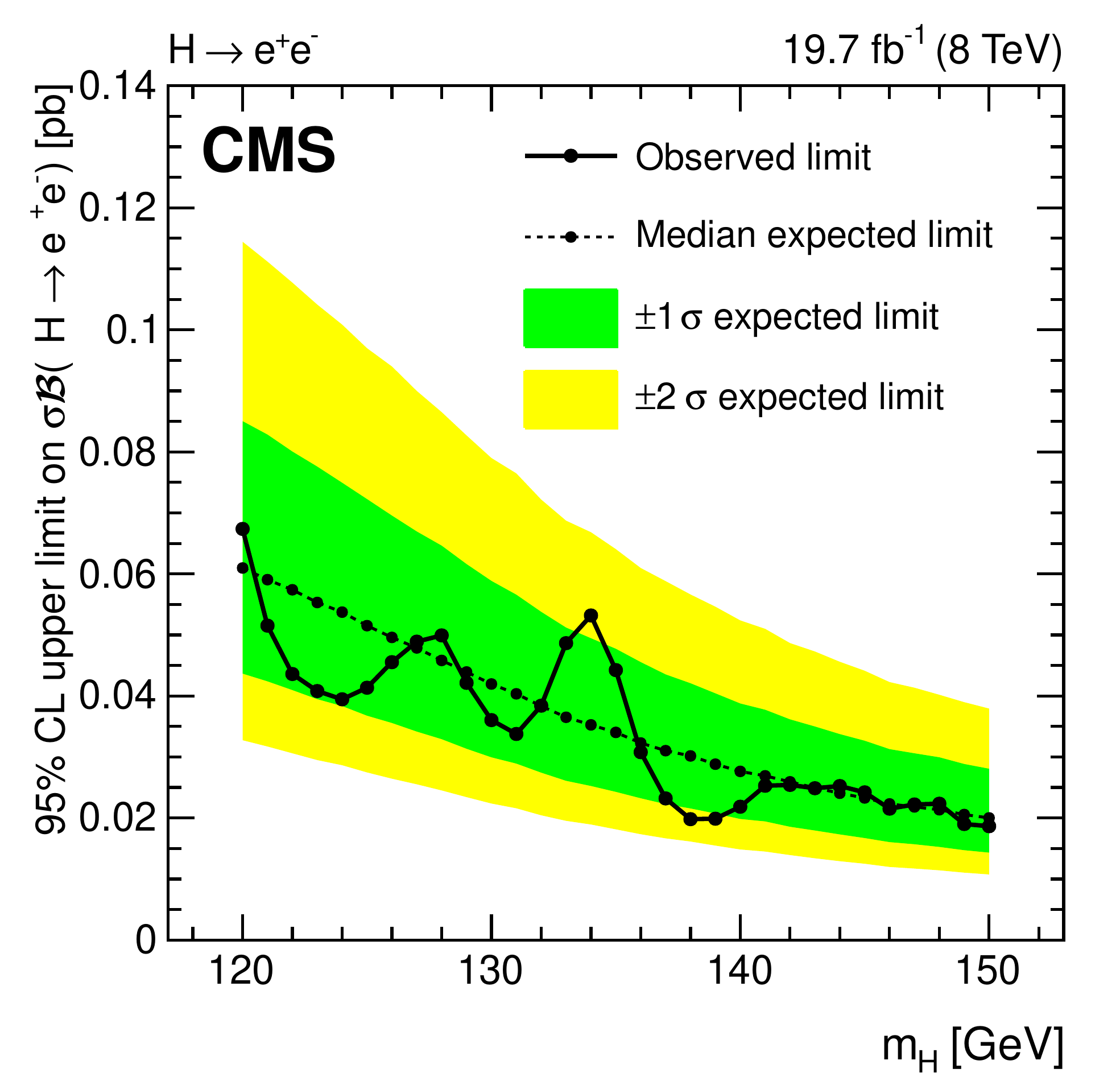}

     \caption{ Mass scan for the background-only expected and observed combined exclusion limits (left). Exclusion limits on $\sigma \BF$ are shown for
      $\hmm$ (middle),
      and for $\hee$ (right), both for 8 $\TeV$.
      Theoretical uncertainties on the
      cross sections and branching fraction are omitted, and the relative contributions
      of GF, VBF, and VH are as predicted in the SM~\cite{Khachatryan:2014aep}.
      }
     \label{fig:Hll}
\end{figure}

For a Higgs boson mass, $\mh$, of 125 $\GeV$, the SM prediction for the Higgs to
$\mumu$ branching fraction, $\BF(\hmm)$,
is among the smallest accessible at the CERN LHC,
$2.2\times 10^{-4}$~\cite{Denner:2011mq},
while the SM prediction for \BF(\hee) of approximately $5\times10^{-9}$
is inaccessible at the LHC.
Experimentally, however, $\hmm$ and $\hee$ are the cleanest of the fermionic decays.
The clean final states allow a better sensitivity, in terms of
cross section, $\sigma$, times branching fraction, $\BF$,
than $\htautau$.
In addition, a measurement of the $\hmm$ decay
probes the Yukawa coupling of the Higgs boson to
second-generation fermions, an important input in understanding the
mechanism of electroweak symmetry breaking in the SM~\cite{Plehn:2001qg,Han:2002gp}.
Deviations from the SM expectation could also be a sign of BSM physics~\cite{Vignaroli:2009vt,Dery:2013rta}.
The \hmm search is performed on data corresponding to integrated luminosities of
$5.0\pm0.1$ $\fbinv$ at a centre-of-mass
energy of 7 $\TeV$ and $19.7\pm0.5$ $\fbinv$ at 8 $\TeV$,
while the $\hee$ search is only performed on the 8 $\TeV$ data~\cite{Khachatryan:2014aep}.
Results are presented for Higgs boson masses between 120 and 150 $\GeV$.
Events are split into categories corresponding to different
production topologies and dilepton invariant mass resolutions.
The signal strength is then extracted using a simultaneous fit to
the dilepton invariant mass spectra in all of the categories.
No significant $\hmm$ signal is observed. Upper limits are set on the signal strength
at the 95\% CL, as shown in the left plot of Fig.~\ref{fig:Hll}.
The combined observed limit on the signal strength,
for a Higgs boson with a mass of 125 $\GeV$,
is 7.4, while
the expected limit is $6.5^{+2.8}_{-1.9}$.
Assuming the SM production cross section, this corresponds to an upper
limit of 0.0016 on $\BF(\hmm)$.
For a Higgs boson mass of 125 $\GeV$, the best fit signal strength
is $0.8^{+3.5}_{-3.4}$.
In the $\hee$ channel, SM Higgs boson decays are far too rare to detect, and no signal is observed.
For a Higgs boson mass of 125 $\GeV$, a 95\% CL upper limit
of 0.041 pb is set on $\sigma  \BF(\hee)$ at 8 $\TeV$, as shown in the right plot of Fig.~\ref{fig:Hll}.
Assuming the SM production cross section, this corresponds to an upper limit
on $\BF(\hee)$ of 0.0019, which is approximately $3.7\times10^5$
times the SM prediction.
For comparison, the $\hmm$ observed 95\% CL upper limit on
$\sigma  \BF(\hmm)$ as shown in the middle plot of Fig.~\ref{fig:Hll} is 0.033 pb (using only 8 $\TeV$ data),
which is 7.0 times the expected SM Higgs boson cross section.

\section{Invisible decays }
\label{Hinv}

A number of models for physics beyond the SM allow for invisible decay modes of the
Higgs boson, such as decays to neutralinos in supersymmetric models~\cite{Belanger:2001am} or
graviscalars in models with extra spatial dimensions~\cite{Giudice:2000av,Dominici:2009pq}.
More generally, invisible Higgs boson decays
can be realised through interactions between the Higgs boson and dark matter (DM)~\cite{Shrock:1982kd}.
Direct searches for invisible decays of the Higgs boson increase the sensitivity to the
invisible Higgs boson width beyond the indirect constraints. The typical signature at the LHC is a large missing
transverse momentum recoiling against a distinctive visible system.
Firstly a combination of searches for invisible decays of the
Higgs boson using data collected during 2011, 2012, and 2015 are presented~\cite{Khachatryan:2016whc}.
The data collected with the CMS detector at the LHC correspond to integrated luminosities of 5.1, 19.7, and 2.3 $\fbinv$ at centre-of-mass
energies of 7, 8, and 13 $\TeV$, respectively.
The combination includes searches targeting Higgs boson production in the ZH mode, in which a Z boson decays to
$\ell^{+}\ell^{-}$ or $\bbbar$, and the qqH mode, which is the most sensitive channel. The combination also
includes the first searches at CMS targeting VH production, in which the vector boson decays hadronically,
and the ggH mode in which the Higgs boson is produced in association with jets. No significant deviations from the SM predictions
are observed and upper limits are placed on the branching fraction for the Higgs boson decay to invisible particles. The combination of all searches
yields an observed (expected) upper limit on $\brinv$ of $\higgsbrobs$ $(\higgsbrexp)$ at the 95\% confidence level, assuming SM production of the Higgs boson.

\begin{figure}[hbt]
  \centering
  \includegraphics[width=0.41\textwidth]{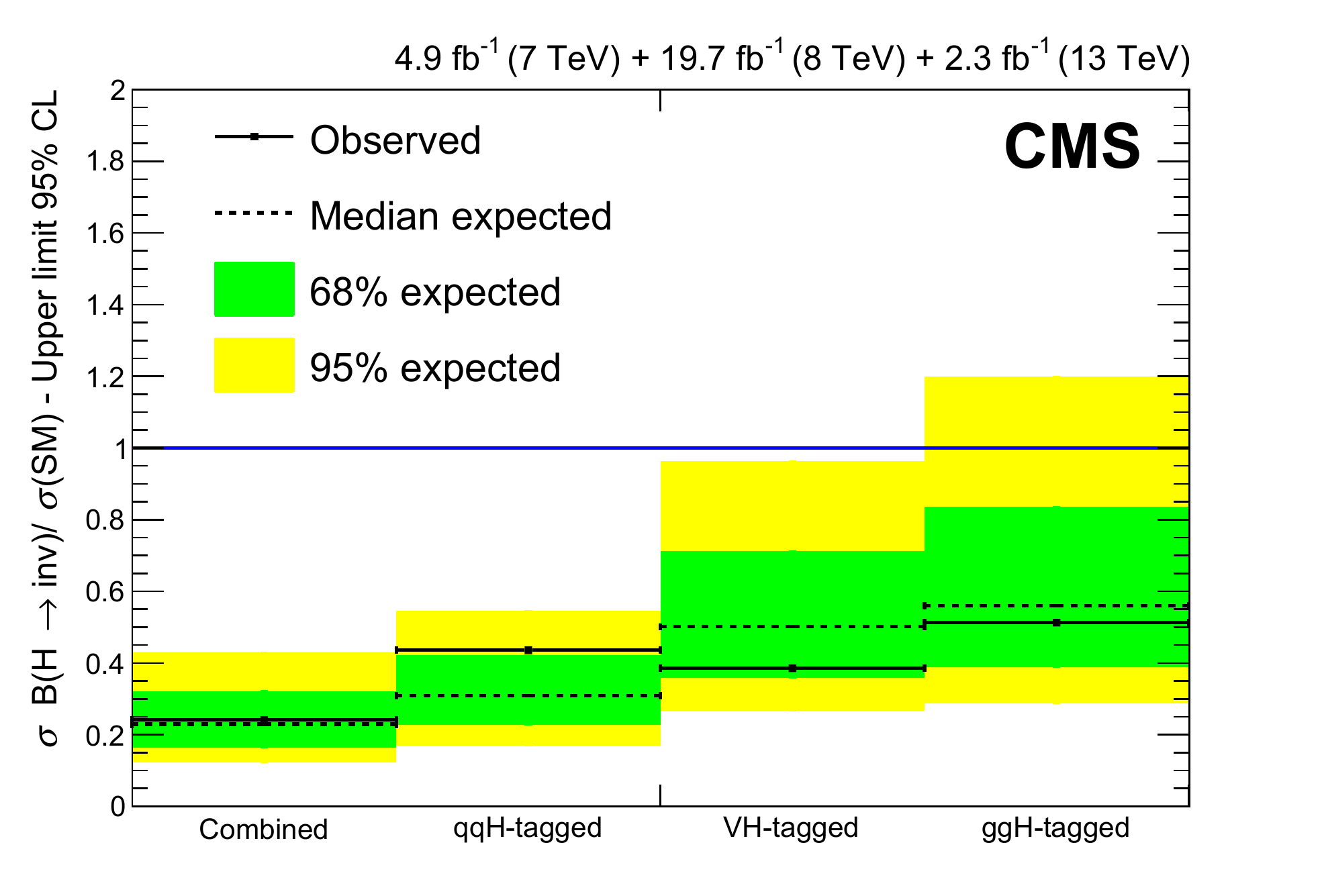}
   \includegraphics[width=0.285\textwidth]{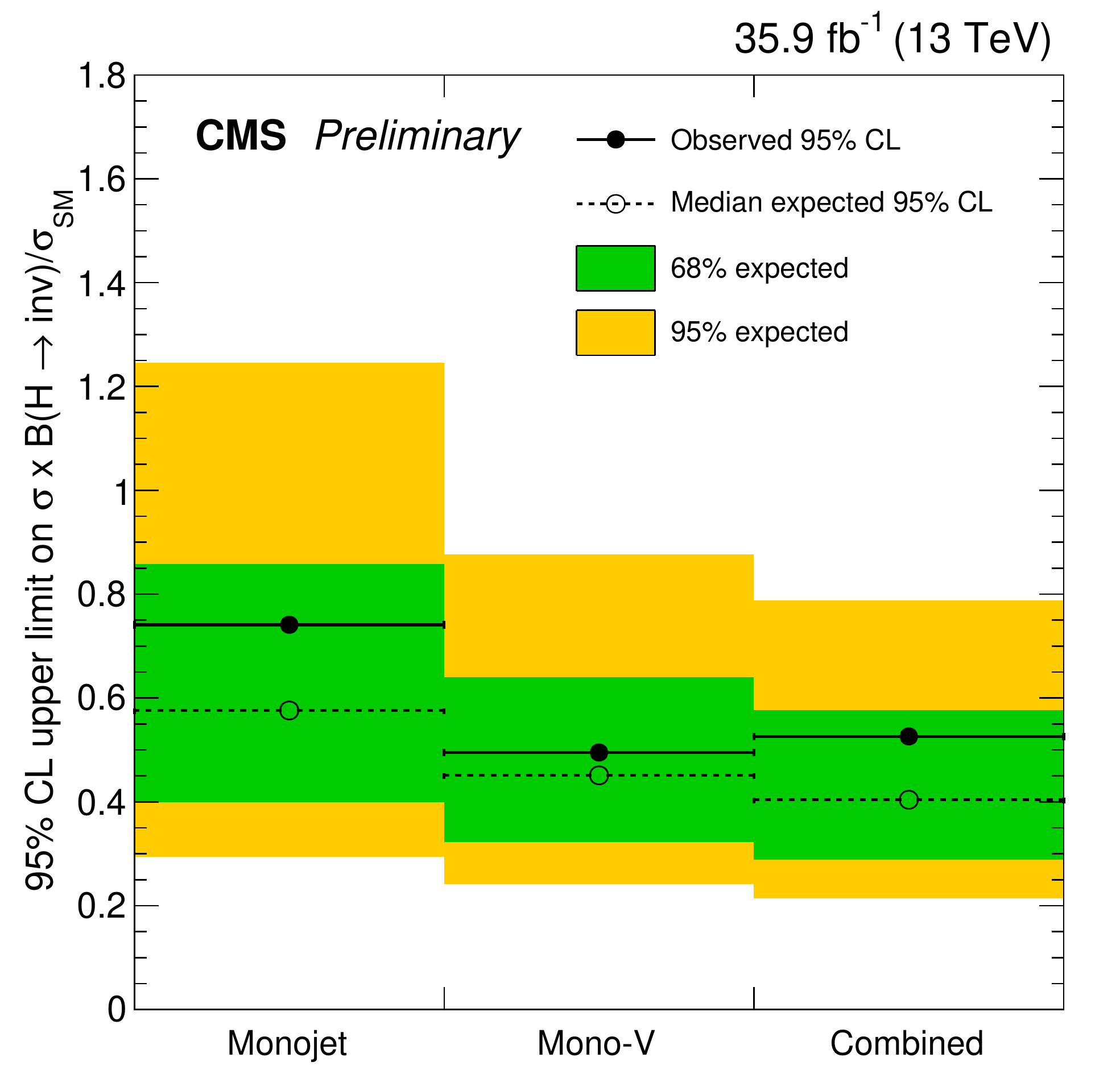}
   \includegraphics[width=0.285\textwidth]{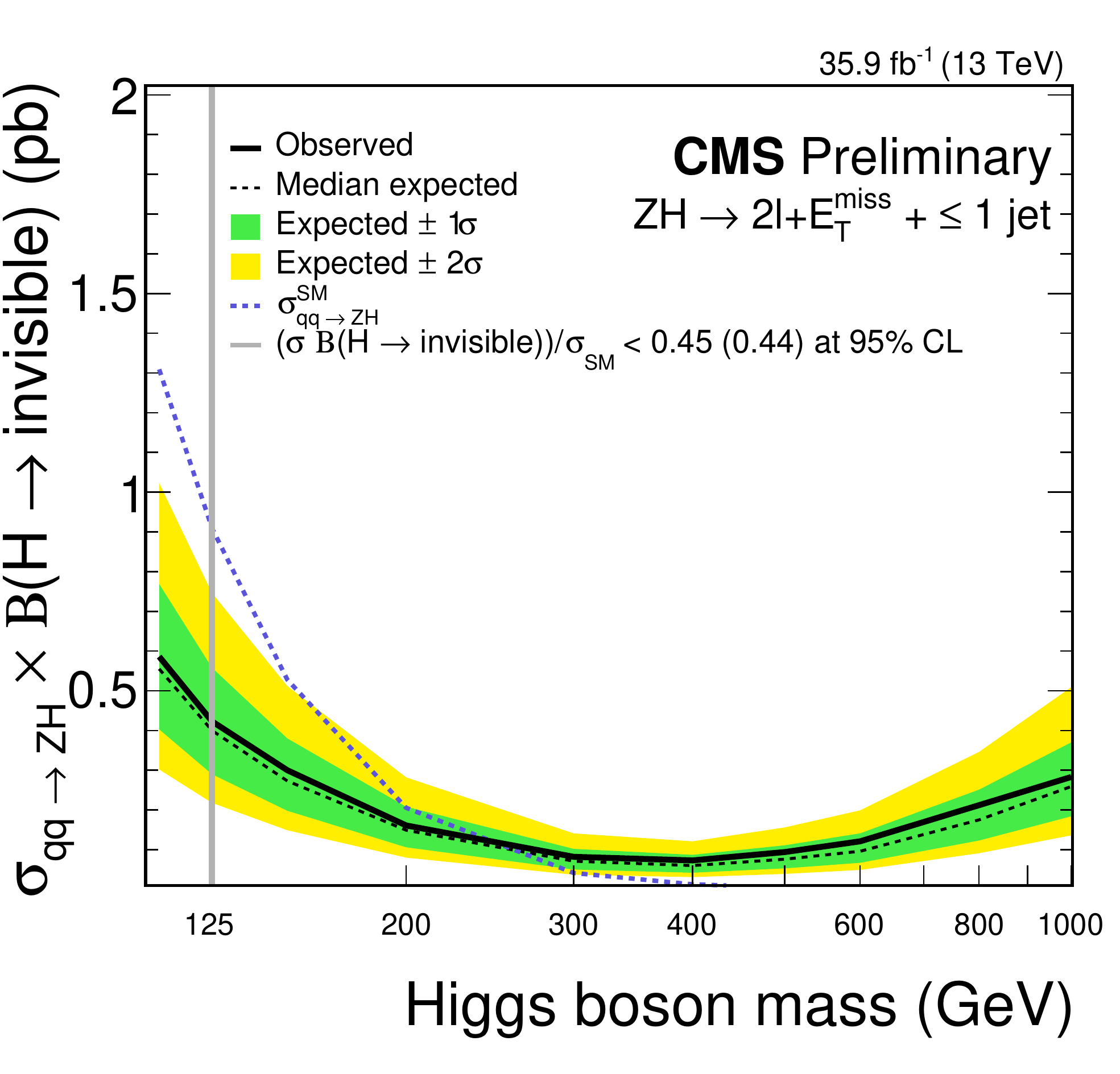}
    \caption{ (Left) Observed and expected 95\% CL limits on $\sigma \times \brinv / \sigma_{\textrm{SM}}$ for individual combinations
    of categories targeting qqH, VH, and ggH production, and the full combination assuming
    a Higgs boson with a mass of 125 $\GeV$~\cite{Khachatryan:2016whc}. (Middle) Expected (dotted black line) and observed (solid black line)
      95\% CL upper limits on the invisible branching fraction of the 125 $\GeV$
      SM-like Higgs boson. Limits are shown for the monojet and mono-V categories
      separately, and also for their combination~\cite{CMS:2017tbk}. (Right) Expected and observed 95\% CL upper
        limits on the production cross section times branching fraction,
 $\sigma_{ZH} \times \brinv$ as a function
        of the Higgs boson mass~\cite{Sirunyan:2017qfc}.
 }
    \label{fig:Hinv}
\end{figure}

Sear for Higgs boson to invisible particles in final states with an energetic jet (Monojet) or a hadronically decaying W or Z boson (Mono-V) based on the 2016 data sample of proton-proton collisions at $\sqrt{s} = 13~\mathrm{TeV}$ corresponding to an integrated luminosity of $35.9~\mathrm{fb}^{-1}$ is also presented~\cite{CMS:2017tbk}. The observed (expected) 95\% CL upper limit on the invisible branching fraction of the
Higgs boson, $\sigma \times \brinv / \sigma_{\textrm{SM}}$,
is found to be 53\% (40\%). The limits are summarized in Fig.~\ref{fig:Hinv} (Middle).
Additionally  search for Higgs boson decaying invisibly and produced in association with the Z boson has been updated based on the 2016 data sample with an integrated luminosity of $35.9~\mathrm{fb}^{-1}$~\cite{Sirunyan:2017qfc}. The 95\% CL median expected and observed upper limits
 on the production cross section times branching fraction,
 $\sigma_{ZH} \times C$, computed with the asymptotic $CL_{s}$
 method are shown in Figure~\ref{fig:Hinv} (Right) for the $\met$-shape analysis.
 Assuming the SM production rate, the 95\% observed (expected) CL upper limit on $\brinv$ is
0.45 (0.44) using the $\met$-shape analysis, and 0.40 (0.42) using the multivariate analysis.
The $\mathrm{gg} \to Z(\ell\ell)\PH$ process has been considered only for the 125 GeV mass point.

\section{Lepton flavour violating decays}
\label{Hlfv}

\begin{figure*}[hbtp]\centering
 \includegraphics[width=0.515\textwidth]{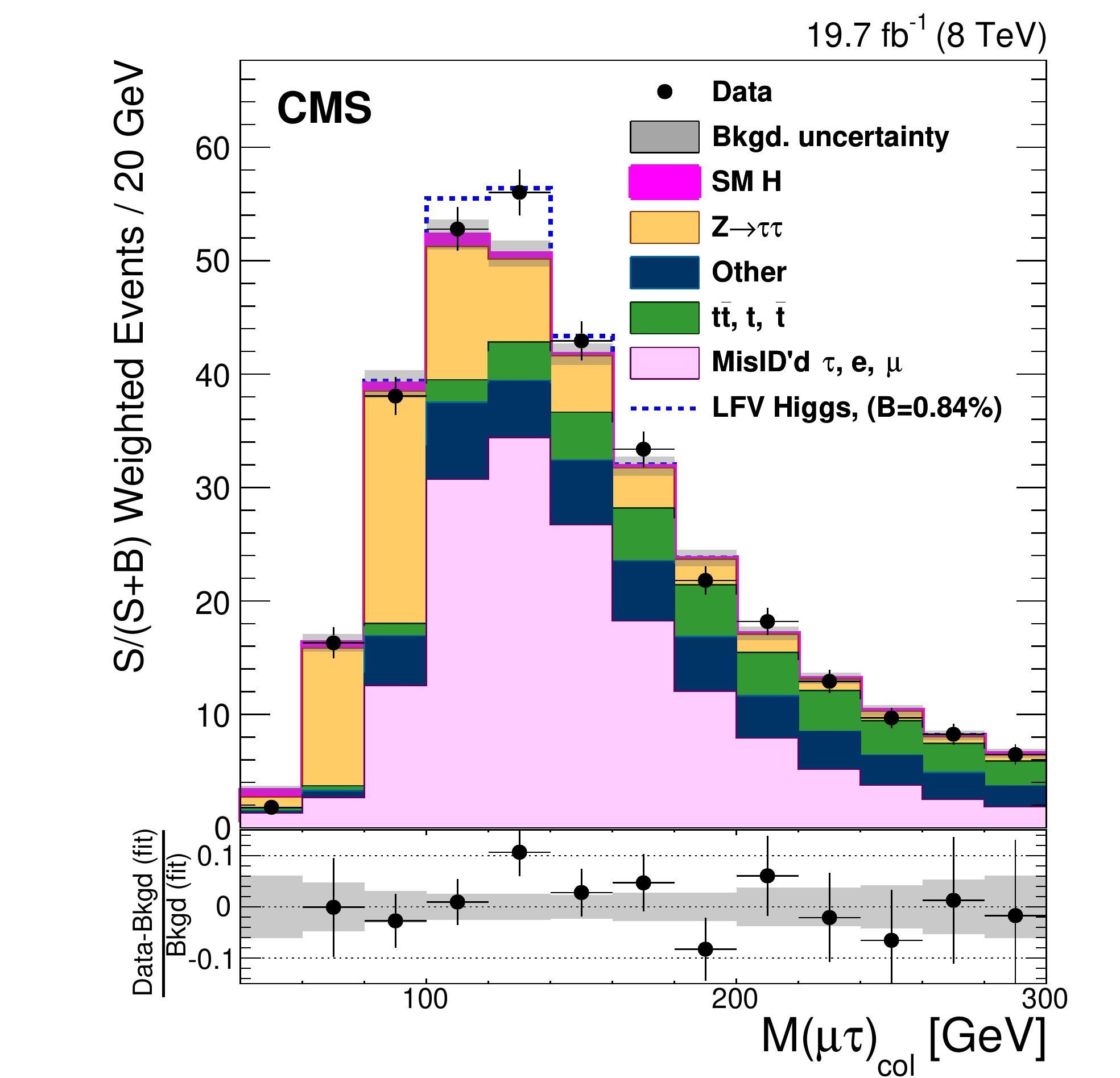}
 \includegraphics[width=0.45\textwidth]{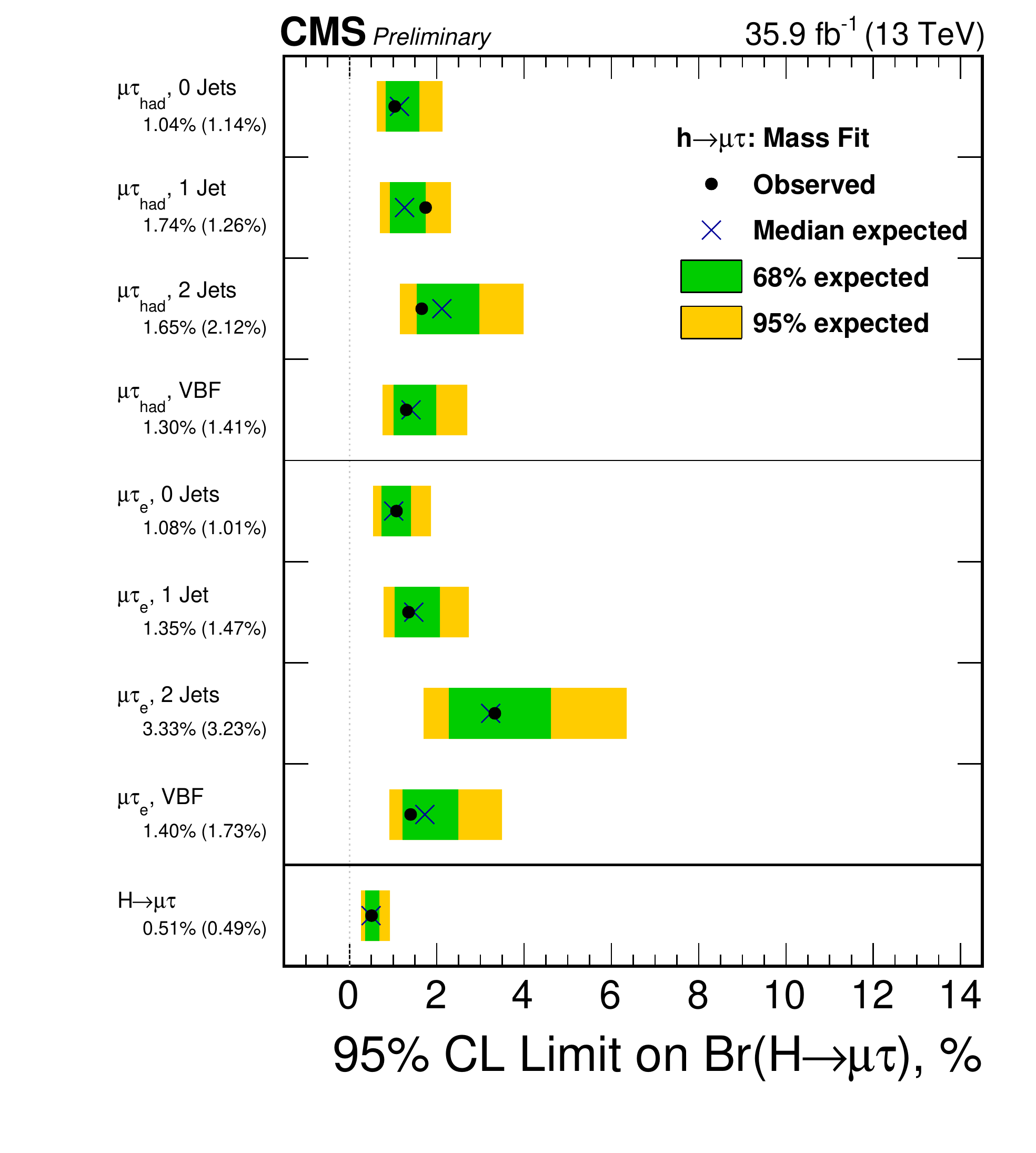}
 \caption{(Left) Distribution of the collinear mass $M_\text{col}$ for all categories combined, with each category weighted by significance ($\mathrm{S}/(\mathrm{S}+\mathrm{B})$)~\cite{Khachatryan:2015kon}. (Right) Observed and expected 95\% CL upper limits on the $\mathcal{B}(\PH \to \Pgm \Pgt)$ for each individual category and combined, from $M_\text{col}$-fit analysis~\cite{CMS:2017onh}.
}
\label{fig:Hmt_mcol_all_global_weighted}\end{figure*}

In the standard model (SM), lepton flavour violating (LFV)  decays of the Higgs boson are
forbidden.  Such decays can  occur naturally in models with more than one Higgs boson
doublet~\cite{Bjorken:1977vt}. Based on $\sqrt{s}=8\TeV$ and 19.7 $\fbinv$ proton-proton collision data,
CMS published the results of the search for a LFV decay of a Higgs boson with $\mh=125$ $\GeV$ in three channels,
$\PH \to \Pgm \Pgt$~\cite{Khachatryan:2015kon}, $\PH \to \Pe \Pgt$ and $\PH \to \Pe \Pgm$~\cite{Khachatryan:2016rke}.
The results from  the $\PH \to \Pgm \Pgt$ channel~\cite{Khachatryan:2015kon} combined the $\PH \to \Pgm \Pgt_{\Pe}$ and $\PH \to \Pgm \tauh$ decays, where  $\Pgt_{\Pe}$ and   $\tauh$ are tau leptons reconstructed in the electronic and hadronic decay channels, respectively.
The results show an excess of data with respect to the SM background-only hypothesis at
$M_{\PH} =125\GeV$ with a significance of $2.4$ standard deviations ($\sigma$). The collinear mass $M_\text{col}$,
which provides an estimator of the reconstructed Hmass using the observed decay products,
is shown in Figure~\ref{fig:Hmt_mcol_all_global_weighted} (Left).
A constraint is set on the branching fraction  $\mathcal{B}(\PH \to \Pgm \Pgt)<1.51\%$
at 95\% confidence level (CL), while the best fit branching fraction is $\mathcal{B}(\PH \to \Pgm \Pgt)=(0.84^{+0.39}_{-0.37})\%$.
Based on the 2016 data sample of proton-proton collisions at $\sqrt{s} = 13~\mathrm{TeV}$ corresponding to an integrated luminosity of $35.9~\mathrm{fb}^{-1}$,
CMS updated the search  for LFV decays of the Higgs boson with $M_{\PH}=125$ $\GeV$. The updated search was performed in four decay channels,  $\PH \to \Pgm \Pgt_{\Pe}$,
$\PH \to \Pgm \tauh$, $\PH \to \Pe \Pgt_{\Pgm}$, $\PH \to \Pe \tauh$, where $\Pgt_{\Pe}$,$\Pgt_{\Pgm}$ and $\tauh$ correspond to the electronic, muonic and hadronic decay channels of $\tau$ leptons, respectively~\cite{CMS:2017onh}. No evidence is found for LFV Higgs boson decays.
The observed~(expected) limits on the branching fraction of the Higgs boson to $\mu\tau$ and to $\Pe\tau$ are found to be
less than 0.25(0.25)\% and 0.61(0.37)\%, respectively.

\begin{figure*}[hbtp]\centering
 \includegraphics[width=0.48\textwidth]{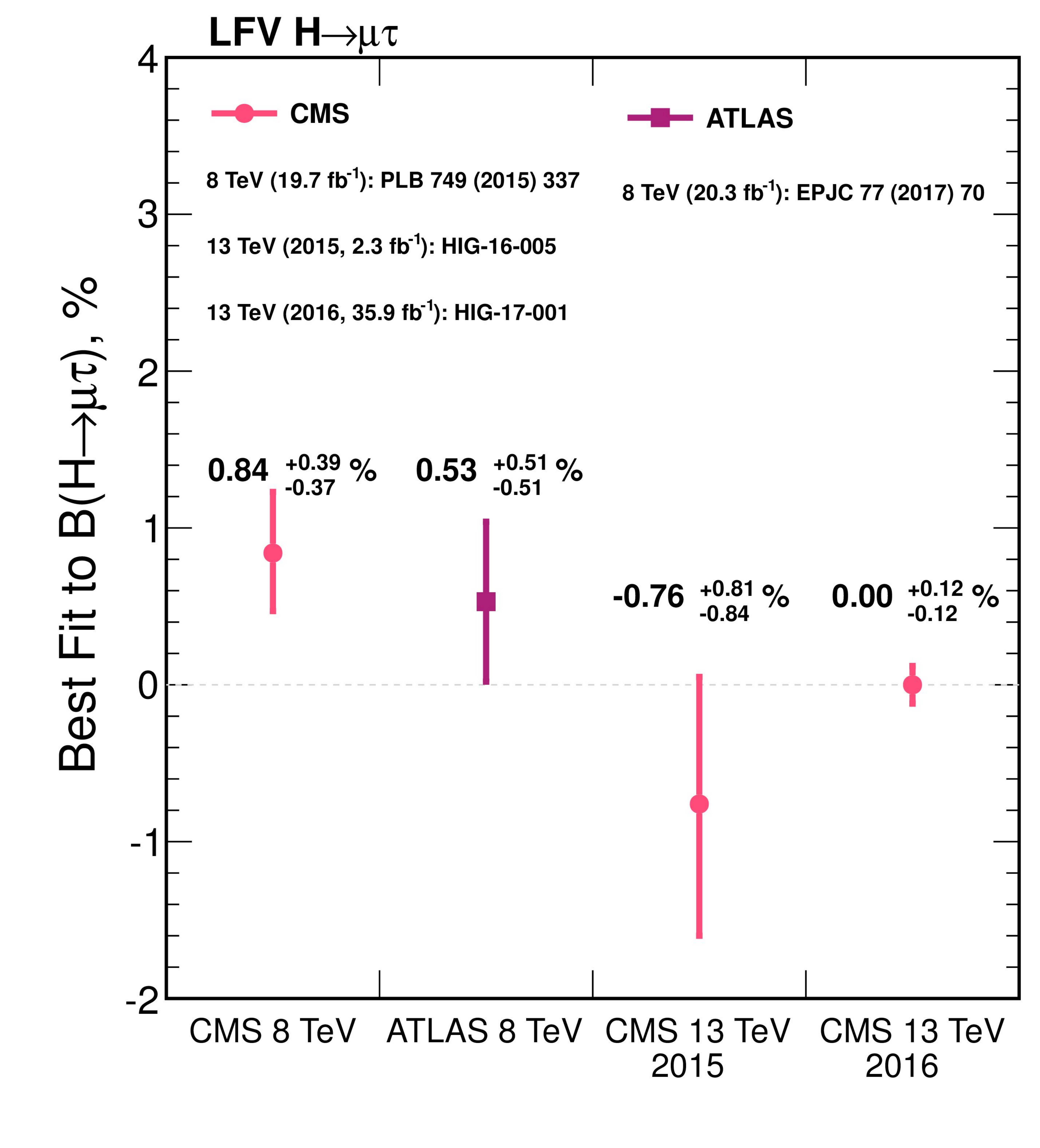}
 \includegraphics[width=0.48\textwidth]{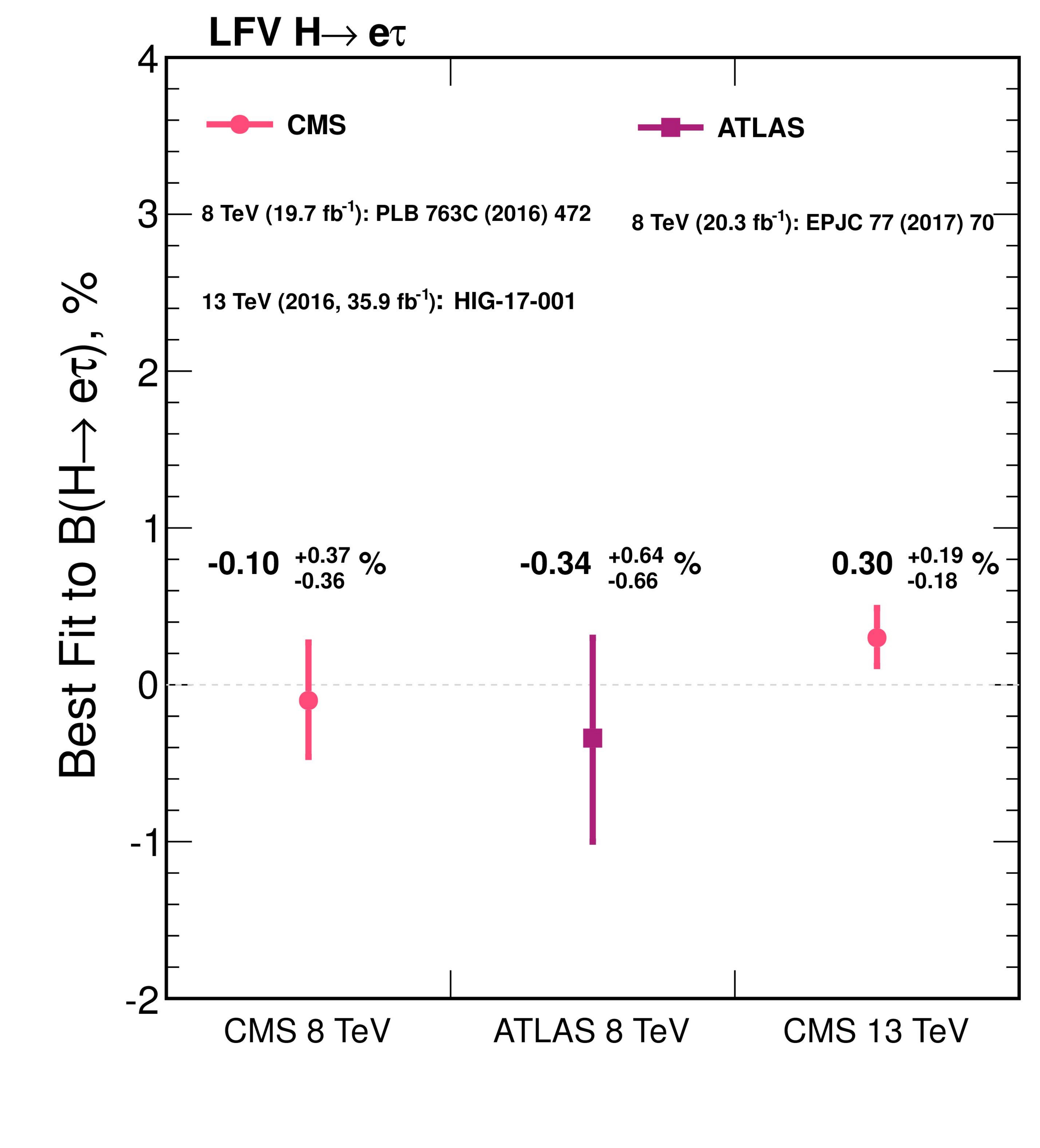}
 \caption{ Summary of the best fit branching fractions of $\PH \to \Pgm \Pgt$ (left) and $\PH \to \Pe \Pgt$ (right) at the LHC.
}
\label{fig:H_LFV_Summary}\end{figure*}

The ATLAS Collaboration reported searches for $\PH \to \Pe \Pgt$ and $\PH \to \Pgm \Pgt$, finding no significant excess of events over the background expectation~\cite{Aad:2016blu,Aad:2015gha}. The best fit branching fractions of $\PH \to \Pe \Pgt$ and $\PH \to \Pgm \Pgt$ obtained from CMS Collaboration and ATLAS Collaboration
are summarized in Figure~\ref{fig:H_LFV_Summary}.

\section{Decay to light scalars or pseudo-scalars}
\label{Haa}

\begin{figure}[htbp!]
\begin{center}
\includegraphics[angle=0,width=0.50\textwidth]{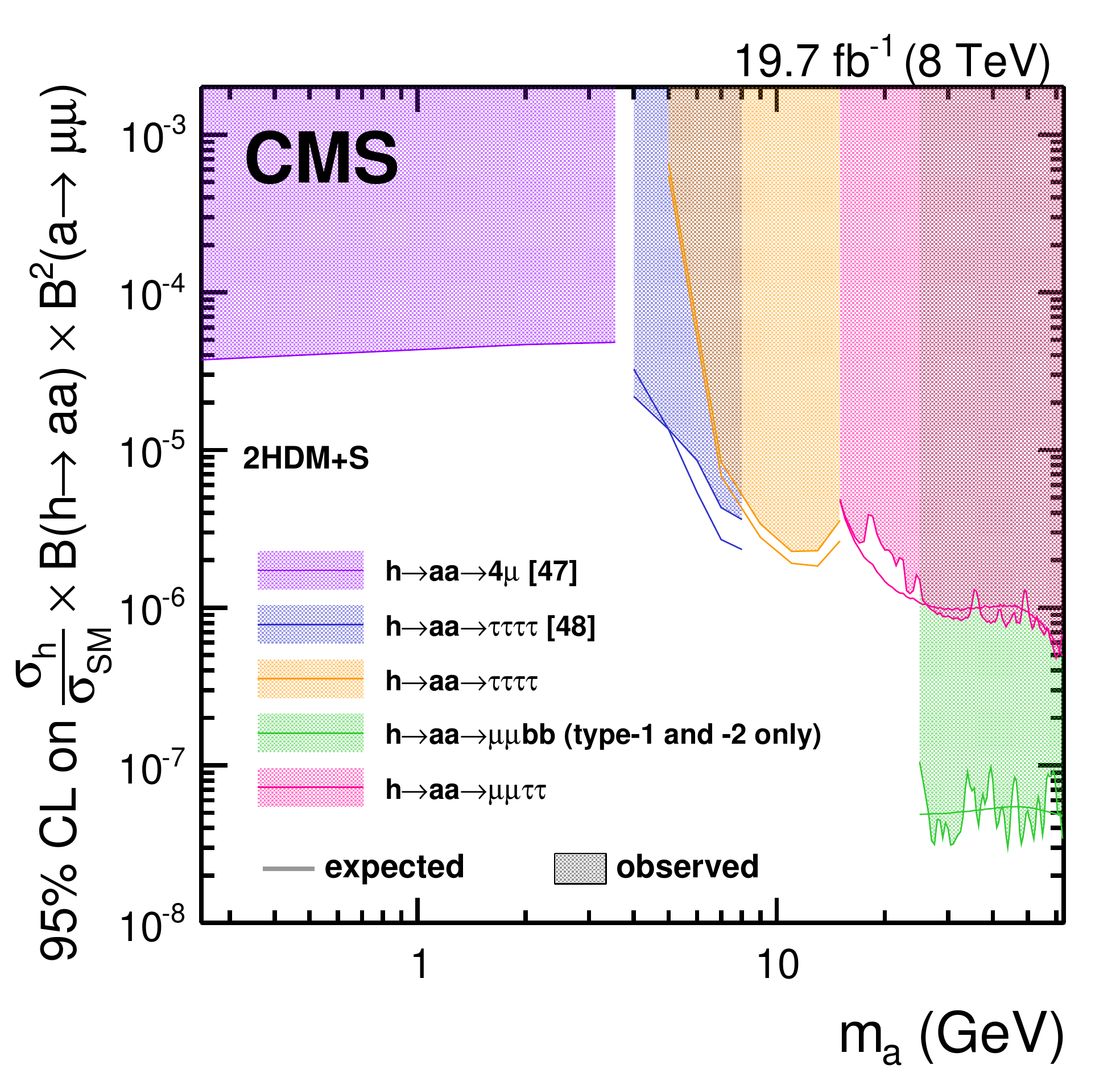}
\caption{Expected and observed 95\% CL exclusion limits on $({\sigma_{\Hb}}/{\sigma_{\textrm{SM}}}) \, \mathcal{B}(\Hb\to\ab\ab)\, \mathcal{B}^2(\ab\to\Pgm^+\Pgm^-)$ for various exotic $\Hb$ boson decay searches performed with data collected at 8 $\TeV$ with the CMS detector, assuming that the branching fractions of
the pseudoscalar boson to muons, $\Pgt$ leptons and \PQb quarks follow the assumption in~\cite{Khachatryan:2017mnf}, which implies that the limit shown for $\processmmbb$ is valid only in type-1 and -2 2HDM+S~\cite{CMS:2016tgd}.
}
\label{fig:compa_modelindependent}
\end{center}
\end{figure}

There are many well-motivated models that predict the existence of decays of the Higgs boson to non-SM particles such as lighter scalars or pseudoscalars~\cite{Dermisek:2006wr,Chang:2008cw,Curtin:2013fra}.
Both 2HDM and NMSSM may contain a light enough pseudoscalar state ($\ab$), which can yield a large $\Hb\to\ab\ab$ branching fraction.
Four searches for decays of the 125 $\GeV$ Higgs boson to pairs of lighter scalars or pseudoscalars are performed at CMS
with $\sqrt{s}=8\TeV$ and 19.7 $\fbinv$ proton-proton collision data~\cite{Khachatryan:2015wka,Khachatryan:2017mnf},
$\processtttt$, $\processmmbb$, $\processmmtt$ and $\processmmmm$, where the symbol $\ab$ refers to both the light scalar and light pseudoscalar for notational simplicity.
The data were found to be compatible with SM predictions. Upper limits are set on the product of the cross section and branching fraction for several signal processes.
Searches for non standard decays of the SM-like Higgs boson to a pair of light pseudoscalar bosons are interpreted in the context of 2HDM+S.
Because $\mathcal{B}(\ab\to\Pgt^{+}\Pgt^{-})$ is directly proportional to $\mathcal{B}(\ab\to\Pgm^+\Pgm^-)$ in any type of 2HDM+S and so is $\mathcal{B}(\ab\to \PQb\overline{\PQb})$ in type-1 and -2, the results of all analyses can be expressed as exclusion limits on $\frac{\sigma(\Hb)}{\sigma_{\textrm{SM}}} \, \mathcal{B}(\Hb\to \ab\ab)\, \mathcal{B}^2(\ab\to\Pgm^+\Pgm^-)$, as illustrated in Fig.~\ref{fig:compa_modelindependent}.
The exact value of $\mathcal{B}(\ab\to\Pgm^+\Pgm^-)$ depends on the type of 2HDM+S,
on $\tan\beta$ and on the pseudoscalar boson mass. No significant excess of events is observed for any of the five analyses.

With $\sqrt{s}=13\TeV$ and 2.8 $\fbinv$ proton-proton collision data during the 2015 data taking campaign,
CMS updated the search for $h \rightarrow 2a + X \rightarrow 4\mu +X$ where $X$ denotes possible additional particles from cascade decays of a Higgs boson~\cite{CMS:2016tgd}, motivated by NMSSM and Dark SUSY models~\cite{ArkaniHamed:2008qn, Baumgart:2009tn, Falkowski:2010cm}. These data are used to search for new light bosons with a mass in the range $0.25-8.5~\mathrm{GeV}/c^2$ decaying into muon pairs. No excess is observed in the data, and a model-independent upper limit on the product of the cross section, branching fraction and acceptance is derived. The results are interpreted in the context of two benchmark models, namely, the next-to-minimal supersymmetric standard model, and dark SUSY models including those predicting a non-negligible light boson lifetime, with the details in~\cite{CMS:2016tgd}.

\section{Summary}
\label{sum}

Searches for rare and exotic decays of the 125 GeV Higgs boson performed with data collected with the CMS experiment have been presented.
The analyzed data correspond to the full LHC run-1 dataset collected during 2011 and 2012 at $\sqrt{s}=7-8~\TeV$ and to the run-2 data at
$\sqrt{s}=13\TeV$ collected during 2015 and 2016.
Rare Higgs decays are extremely sensitive to new physics if additional Higgs couplings exist. Many rare decays have not been observed yet, but may become observable in the next few years. Exotic Higgs decays would bring direct evidence of such new physics. Tight limits on $\PH \to \Pe \Pgt$ and $\PH \to \Pgm \Pgt$ have been set by CMS using data collected in 2016 and the $2.4\sigma$ excess observed in run-1 with $\PH \to \Pgm \Pgt$ decay has been ruled out. No hint for new physics has been found when looking for Higgs invisible decays and $\Hb\to\ab\ab$ searches. The results are interpreted according to beyond-SM (BSM) Higgs scenarios, which include the Two Higgs Doublet Model (2HDM) and the Singlet Model. Stringent limits have been set on the existence of such processes.

\Acknowledgements

I would like to thank the ICFNP2017 organizers for their hospitality and
the wonderful working environment. I acknowledge the support from
National Natural Science Foundation of China (No. 11505208 and No. 1161101027),
China Ministry of Science and Technology (No. 2013CB838700).

\end{document}